\documentclass[aoas]{imsart}
\RequirePackage{natbib}

\usepackage{tabu}
\usepackage{adjustbox}
\usepackage{caption}
\usepackage{subcaption}

\usepackage{multirow}
\usepackage{algorithm,algorithmic,enumerate}
\usepackage{amsmath}
\usepackage{mathtools}
\usepackage{amssymb}
\usepackage{amsfonts}
\usepackage{verbatim}
\usepackage{graphicx}
\usepackage{array}
\usepackage{times,hyperref}
\usepackage{graphicx,color}
\usepackage{bm}
\usepackage{natbib}

\newcommand{ \bmm }[1]{ \mbox{\boldmath ${#1}$} }
\def\Argmax{\mathop{\rm arg\,max}\limits}
\def\Argmin{\mathop{\rm arg\,min}\limits}

\newcommand{\ba}{\bmm{\alpha}}
\newcommand{\bxi}{{\bf X}_i}
\newcommand{\bzi}{{\bf Z}_i}
\newcommand{\bri}{{\bf R}_i}

\newcommand{\bwi}{{\bf W}_i^{(t)}}

\newcommand{\bi}{\bmm{b}_i}

\newcommand{\ymis}{\bmm{y}_\textrm{mis}}
\newcommand{\yobs}{\bmm{y}_\textrm{obs}}
\newcommand{\bO}{{\bf \Omega}}
\newcommand{\Ot}{{\bf \Omega}^{(t)}}
\newcommand{\E}{\textrm{E}}
\newcommand{\var}{\textrm{var}}

\newcommand{\pr}{\textrm{Pr}}

\def\bSig\mathbf{\Sigma}

\usepackage[figuresright]{rotating}

\arxiv{arXiv:0000.0000}

\begin{document}
\begin{frontmatter}
\title{A mixed-effects model for incomplete data with Batch-level Abundance-Dependent Missing-data Mechanism}
\runtitle{A mixed-effects model for data with batch-level non-ignorable missingness}

\begin{aug}
\author{ \snm{Lin S. Chen$^1$}\ead[label=e1]{lchen@health.bsd.uchicago.edu}},
\author{ \snm{Jiebiao Wang$^1$}\ead[label=e2]{jwang88@uchicago.edu}},
\author{ \snm{Xianlong Wang$^2$}\ead[label=e3]{xwan2@fhcrc.org}}
\and
\author{ \snm{Pei Wang$^3$}\ead[label=e4]{pei.wang@mssm.edu}}

\runauthor{Chen et al.}

\affiliation{$^1$University of Chicago, $^2$Fred Hutchinson Cancer Research Center and $^3$Icahn School of Medicine at Mount Sinai}

\address{
L. S. Chen\\
J. Wang\\
Department of Public Health Sciences\\
University of Chicago\\
5841 S Maryland Ave\\
Chicago, Illinois, USA.\\
\printead{e1}\\
\phantom{E-mail:\ }\printead*{e2}}

\address{
X. Wang\\
Division of Public Health Sciences\\
Fred Hutchinson Cancer Research Center\\
University of Pennsylvania\\
1100 Fairview Ave N
Seattle, Washington, USA.\\
\printead{e3}\\
}
\address{
P. Wang\\
Icahn Institute of Genomics and Multiscale Biology\\
Icahn School of Medicine at Mount Sinai\\
1470 Madison Ave, S8-102\\
New York, New York, USA\\
\printead{e4}\\
}

\end{aug}

\begin{abstract}
In mass spectrometry- (MS-) based quantitative proteomics research, the emerging iTRAQ (isobaric tag for relative and absolute quantitation) technique has been widely adopted for high throughput protein profiling. In a typical iTRAQ proteomics study, samples are grouped into batches and each batch is processed by one iTRAQ multiplex experiment, in which the abundances of thousands of proteins/peptides in a batch of samples can be measured simultaneously. The iTRAQ technique greatly enhances the throughput of protein quantification. However, the technical variation across different iTRAQ multiplex experiments is often large due to the dynamic nature of MS instruments. This leads to strong batch effects in the iTRAQ data. Moreover, the iTRAQ data often contain substantial batch-level non-ignorable missingness.
Specifically,
the abundance measures of a given protein/peptide are often either observed or missing altogether in all the samples from the same batch, with the missing probability depending on the combined batch-level abundances.
We term this unique missing-data mechanism as the Batch-level Abundance-Dependent Missing-data mechanism (BADMM). We introduce a new method --- \texttt{mixEMM} --- for analyzing iTRAQ data with batch effects and batch-level non-ignorable missingness. The \texttt{mixEMM} method employs a linear mixed-effects model and explicitly models the batch effects and the BADMM in the likelihood function.  With simulation studies, we showed that compared with existing approaches that utilize relative abundances and ignore the missing batches under the missing-completely-at-random assumption, the  \texttt{mixEMM} method achieves more accurate parameter estimation and inference. We  applied the method to an iTRAQ proteomics data from a breast cancer study and identified phosphopeptides differentially expressed between different breast cancer subtypes.  The method can be applied to general clustered data with cluster-level non-ignorable missing-data mechanisms.
\end{abstract}

\begin{keyword}
\kwd{mixed-effects models}
\kwd{the Expectation-conditional-maximization (\texttt{ECM}) algorithm}
\kwd{Batch-level Abundance-Dependent Missing-data Mechanism (BADMM)}
\end{keyword}

\end{frontmatter}

\section{Introduction}\label{s:intro}
\subsection{Quantitative proteomics research and the iTRAQ technique}
Proteins are complex macromolecules responsible for nearly every task of cellular life and essential for the structure, function, and regulation of human tissues and organs. Nonetheless, the discovery of protein biomarkers in cancer diagnosis, prevention and treatment has only achieved modest successes, partially because that the abundances of proteins are difficult to be quantified. To date, MS-based platforms still serve as the workhorses in quantitative proteomics research. Traditional shotgun MS experiments usually process samples one by one; and the process of each sample involves extensive fractionation, resulting in weeks of MS time. The huge time and cost required for such experiments greatly limit the scale of most proteomics studies.

To improve the efficiency of MS-based protein quantification, the iTRAQ (isobaric Tag for relative and Absolute Quantitation) technique was introduced about a decade ago \citep{Ross04,Wiese07}. It enables  multiplexing (comparing) of up to 4 or 8 different samples in one MS-based experiment with 4 or 8 ``channels'' (i.e. 4-plex or 8-plex reagents). Specifically, in an iTRAQ-MS based study, samples are firstly grouped into batches (4 or 8 samples per batch), and then each batch is processed by one iTRAQ multiplex experiment following three steps: (1) intact proteins of each sample are enzymatically digested into smaller segments of amino acid sequences, i.e. peptides; (2) peptides from different samples in one batch are labelled with different isotope-coded covalent tags and are mixed together; (3) the mixtures are introduced into MS instruments, where peptides from different samples in the batch are identified and quantified together. 
In this way, multiple samples can be processed together and that greatly reduces the overall quantification time and cost.
Moreover, iTRAQ labeling was reported to be superior to other competing platforms in quantitative proteomics research \citep{Mertins12}.  In 2014 alone, there are more than 2,000 publications involving iTRAQ experiments according to the Google Scholar. Apparently, the enhancement of throughput with iTRAQ had greatly advanced proteomics research.

\subsection{A motivating iTRAQ proteomics data from the CPTAC project}
In order to improve our ability to diagnose, treat and prevent cancer, the National Cancer Institute launched the Clinical Proteomic Tumor Analysis Consortium (CPTAC, http://proteomics.cancer.gov) to systematically identify proteins that are derived from alterations in cancer genomes \citep{CPTAC,CPTAC2}. The CPTAC has recently conducted global proteome and phosphoproteome profiling of a subset of breast, colon and ovarian cancer samples that have been extensively characterized in  The Cancer Genome Atlas (TCGA , http://cancergenome.nih.gov) \citep{tcga12}. This is so far the first attempt to characterize protein activities in cancer samples using sophisticated proteomics experiments on a large scale. For example, in the breast cancer project, a total of 108 breast cancer tumor samples were analyzed with iTRAQ  experiments, with the goal of identifying proteins related to breast cancer clinical variables and outcomes.

Anther aim of the CPTAC project is to ``set standards, establish procedures, and provide reagents to enable cancer researchers to effectively and reproducibly use proteomics approaches" \citep{CPTAC,CPTAC2}. Advances in methods and tools, especially the ones accounting for the unique characteristics of iTRAQ data, such as the method proposed in this paper, will better facilitate the achievement of those missions, and will in turn lead to improved diagnostics, therapies, and potentially preventive measures for cancer.

In this paper, we will focus on analyzing the phospho-proteomics data from the breast cancer CPTAC study. Phosphorylation is a key post-translational modification and plays a central role in many biological processes. Phosphorylation at different sites of one protein could induce different biological activities. Our goal is to identify individual phosphorylated peptides, i.e., phosphopeptide, up or down-regulated in triple negative breast cancer tumors compared to other subtypes of breast cancer. The investigation will provide important insights into breast cancer etiology and help to identify protein biomarkers.

\subsection{Batch effect and batch-level non-ignorable missing data}

Given the popularity and the efficiency of the iTRAQ technique, there is a pressing need for tailored methods development for iTRAQ data. Though the iTRAQ-based batch-processing greatly reduces the cost and improves the efficiency of data generation, the consequent batch-effects are substantial due to the dynamic nature of MS instrument. To alleviate this problem, a general practice is to include a common reference sample in each batch for quality control. For example, in the 4-plex iTRAQ experiments of the CPTAC breast cancer study, each batch consisted of 3 breast tumor samples and a (same) reference sample. The reference sample was created by combining 40 tumor samples in the CPTAC breast cancer study. Conventional data analyses are usually performed based on the relative abundances of proteins/peptides in the target samples relative to the reference sample in the same batch \citep{Mertins12,Karp10}. This strategy helps to account for the variation across different iTRAQ multiplex experiments to a certain extent. However, due to the complicated process of protein/peptide identification and quantification in the MS instruments, there is a large variation among the measurements of the reference sample across different experiments/batches, and the target samples and the reference sample could be subjected to different variances \citep{Karp10}. The relative abundance measures can not fully capture these data features.

\begin{figure}[h]
\begin{center}
\includegraphics[scale=0.5]{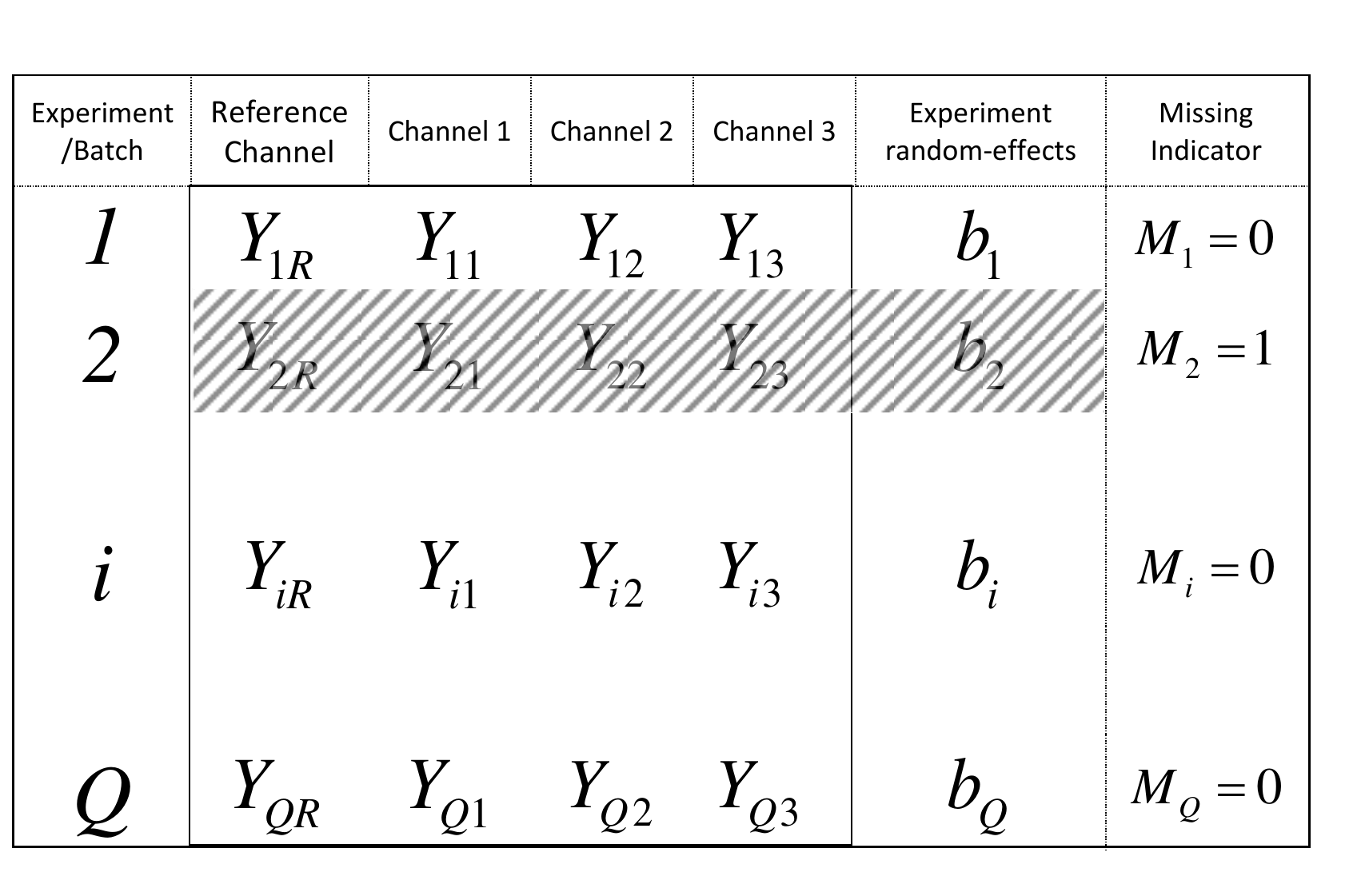}
\end{center}
\caption{An illustration of a 4-plex iTRAQ data for one peptide. There are often tens of thousands of peptides being quantified simultaneously. Let ${\bf Y}_{Q\times 4}$ be the abundance data for the peptide. A total of $3\times Q$ tumor samples are randomly grouped into $Q$ batches and are processed by $Q$ iTRAQ experiments. In each iTRAQ experiment $i$ ($i=1,\ldots,Q$), the three tumor samples are placed on three channels, and a reference sample is also placed on one channel for quality control purposes. The samples in the same batch are processed together, and often will be  observed or missing together. If missing, the missing indicator for the $i$-th batch, $M_i$ is set to be 1. The missing probability of the batch relates to the total peptide abundance level in the batch. The lower the total abundance, the more likely the peptide will be missing in the experiment (batch).  }\label{fig:01}
\end{figure}

Another unique challenge in the iTRAQ data analysis is the substantial amount of batch-level non-ignorable missing data.
It is well known that in the general MS experiments, the lower the abundance of a given peptide is, the more likely the peptide is missing in the output data \citep{Wang06,CPW14}. With iTRAQ-MS experiments, since all of the samples in a batch are processed together, a given peptide is either detected and quantified or missing simultaneously in the samples from the same batch. The missing data probability of the peptide largely depends on the combined abundances of the peptide from all of the batch samples in the experiment (the batch-level abundance). We term this missing-data mechanism in the iTRAQ proteomics data as the ``Batch-level Abundance-Dependent Missing-data mechanism (BADMM)". Figure \ref{fig:01} shows an illustration of the iTRAQ data on one peptide and its BADMM. Subsequently, protein quantification is often obtained as a summary of the peptide abundances in the protein and is also subject to the BADMM.
In addition to BADMM, there may be sporadic missingness created at the individual sample level. Here sporadic missingness refers to the scenario where a peptide/protein is missing in some but not all of the samples from the same batch. Since the proportion of sporadic missing data is usually small (e.g. $<1\%$ sporadic versus $>99\%$ batch-level missingness in the motivating CPTAC data set), we assume these sporadic missing-values are missing-completely-at-random and are ignorable \citep{Rubin76}.\\

Given the presence of severe batch effects, substantial batch level missingness ($\sim 40\%$ in our motivating CPTAC data), and small sample sizes in most iTRAQ proteomics data, it is very essential to account for the batch design and the non-ignorable missingess in a deliberate way to improve the precision of estimation and inference with iTRAQ data. In this work, we propose to directly model the absolute abundances of proteins/peptides and their variance structures considering the batch design.  By modeling the absolute abundances instead of the relative abundances, we can better characterize the variance of protein abundances in target samples, and  improve the power of statistical tests. This strategy has been employed for analyzing other types of proteomics data from targeted mass spectrometry experiments \citep{Chang2012}. Since samples in the same batch are subject to the same experimental conditions and procedures, a mixed-effects model with a random effect for each batch is a natural way to account for the experimental design \citep{LW82}.

The BADMM in the iTRAQ data hinders the direct application of a mixed-effects model. With BADMM, the probability of a protein/peptide being missing in a batch depends on the combined abundance of the protein/peptide in the batch.  The missing data are not-missing-at-random and are non-ignorable \citep{Rubin76}. In order to obtain unbiased estimation and valid inference, the missing-data mechanism needs to be properly modelled and accounted for. Existing work on modelling the non-ignorable missingness in iTRAQ data \citep{Hill08,Luo09,Oberg08} and the selection model for longitudinal data with non-ignorable missingness \citep{Ibrahim90} consider the probability of missingness for a protein/peptide in each sample independently. If a protein/peptide is missing in the entire batch, the values in the batch will be ignored by existing methods, leading to biased estimation and unfaithful inference. In contrast, we propose to model the batch-level missing-data pattern (BADMM) and incorporate it in a mixed-effects model: 
we model the probability of a protein/peptide being missing (in all of the samples) in a batch as a function of the total protein/peptide abundances in the current batch. This probabilistic missing-data mechanism provides an attractive way to account for the characteristics of iTRAQ and MS experimental complexities. Compared to a censoring model \citep{LR02}, it does not depend on a fixed detection threshold and is more flexible, and it better depicts the experimental procedure.

\subsection{Outline}

To properly analyze iTRAQ data, as characterized by the data from the CPTAC project, we introduce \texttt{mixEMM} --- a mixed-effects model coupled with the probabilistic BADMM in Section 2. In Section 3, we use an Expectation and Conditional Maximization (ECM) algorithm to estimate the fixed and random effects in \texttt{mixEMM}. We also present an alternative probability function for BADMM that may be suitable for more general settings. In Section 4, we perform simulations to evaluate the performance of the \texttt{mixEMM} method. In Section 5, we apply the proposed method to the motivating CPTAC iTRAQ data and identify phosphopeptides related to breast cancer subtypes. In Section 6, we summarize the work as a useful tool for analyzing iTRAQ proteomics data, and moreover, as a general framework to handle cluster-level non-ignorable missing-data patterns for data with repeated or clustered measures.

\section{A mixed-effects model with batch-level non-ignorable missingness}
Considering one feature (e.g., a phosphopeptide) of interest,  let $\bmm{Y}=\{\bmm{y}_i\}_{i=1}^Q$ denote the underlying complete (observed and missing) abundances for this feature in all of the samples in the $Q$ batches ($Q=36$ in the motivating CPTAC breast cancer dataset). Specifically, $\bmm{y}_i$ is a $p_i\times 1$ data vector, where $p_i$ is the number of samples in the $i^{th}$ batch.
Suppose this feature is only observed in $Q_{obs}$ batches $(Q_{obs}\leq Q)$. Let $\yobs$ and $\ymis$ denote the observed and missing data, respectively, and ${\bf Y }= \{ \yobs, \ymis \}$.

Samples in the same batch are processed by one iTRAQ multiplex experiment, and are subject to the same experimental procedure and are correlated. We use a linear mixed-effects model to account for such correlations:
\begin{equation} \label{eq:01}
\bmm{y}_i=\bxi \bmm{\alpha}+\bzi \bmm{b}_i+\bmm{e}_i,
\end{equation}
where $\bxi$ is a known fixed design matrix with dimension $p_i\times k$; $\bmm{\alpha}$ is a $k\times 1$ vector of parameters for fixed effects; $\ \bzi$ is a known covariate matrix for random effects with dimension $p_i \times h$; 
 $\bmm{b}_i\sim N(\bmm{0}, {\bf D}_{h\times h})$ represents the random effect coefficient specific to each batch of samples; and $\bmm{e}_i \overset{i.i.d.}{\sim } N(\bmm{0}, {\bf R}_i)$ has diagonal covariance matrix of dimension $p_i\times p_i$. In our data application, $\bxi$ consists of a column of $1$'s, an indicator variable for the reference sample, and a set of clinical variables (for example, cancer subtype indicators); random effect $\bmm{b}_i$ is of length 1 ($h=1$); $\ \bzi$ is a vector of $1$'s; and for 4-plex iTRAQ experiments, ${\bf R}_i$ has diagonal elements $\{\sigma_0^2, \sigma^2,\sigma^2,\sigma^2 \}$ , where $\sigma_0^2$ is the variance corresponding to the reference sample and $\sigma^2$ is the variance of the other three samples. Since the reference sample was created by combining 40 tumor samples in the CPTAC breast cancer study, we expect it to have a different variance than other individual tumor samples.

According to (\ref{eq:01}), we have $\bmm{y}_i\sim N(\bxi \bmm{\alpha}, {\bf\Sigma}_i)$, where ${\bf\Sigma}_i=\bzi{\bf D}\bzi^T+{\bf R}_i$. Our goals are to obtain the maximum likelihood estimates (MLEs) of the fixed and random effects while accounting for the non-ignorable BADMM and to draw inferences on the fixed effects for identifying features related to clinical variables ($\bxi$).

As described in the previous section, for a given feature, the lower its combined abundance across all of the samples in one batch, the more likely all measures of the feature in the batch will be missing during the experiment. Let $M_i$ be the missing indicator of this feature in the $i^{th}$ batch, and $M_i=1$ if the $i^{th}$ batch is missing, otherwise $M_i=0$. We model this BADMM using an exponential probabilistic model:

\begin{equation} \label{eq:02}
\pr(M_i=1|\bmm{y}_i)= g(\bmm{1}^T \bmm{y}_i; \gamma_0, \gamma)=\exp(- \gamma_0- \gamma /p_i \cdot \bmm{1}^T \bmm{y}_i),
\end{equation}
where $\gamma_0$ and $\gamma$ are non negative parameters. Since $\bmm{y}_i$ are abundance measures and are all positive, $\bmm{y}_i>0$, the above probability function always takes value between 0 and 1.

We first treat $\gamma_0$ and $\gamma$ as known missing-data mechanism parameters. We discuss extensions to cases where those parameters are unknown in Section 3.3. Moreover, in Section 3.5, we discuss other flexible probability functions for BADMM.

Our goal is to obtain the MLEs that maximize the observed-data likelihood function considering the missing-data mechanism:
\begin{eqnarray*} 
{\bf \hat \Omega} = \Argmax_\Omega L(\yobs,  {\bf M};{\bf \Omega})
\end{eqnarray*}
where ${\bf \Omega}=\{\bmm{\alpha},\sigma_0^2,\sigma^2, {\bf D}\}$ denotes the set of parameters of interest.
Directly solving the above likelihood function for MLEs is difficult. Therefore, we propose to employ an \texttt{ECM} algorithm, and we term the proposed method as \texttt{mixEMM} (Mixed-Effects Models with BADMM).

\section{An \texttt{ECM} algorithm to calculate MLEs}
If $\ymis$ and $\bmm{b}_i$ were observed, the MLEs for ${\bf R}_i$, $\bf D$ and $\bmm{\alpha}$ based on the likelihood of the complete data $(\yobs, \ymis, \bmm{b}, {\bf M})$ can be easily calculated. Thus, we employ an \texttt{ECM} algorithm \citep{MR93}: in the expectation (E) step of the $(t+1)$-th iteration, we calculate $Q(\bO|\Ot)$ --- the expected value of the log-likelihood given the observed data and current parameter estimates.
In the conditional maximization (CM) step, we obtain the current parameter estimates $\hat\bO^{(t+1)}$ by maximizing $Q(\bO|\Ot)$. Given the proposed BADMM in equation~(\ref{eq:02}), closed form solutions are available in the CM step.  By iterating through the E and  CM steps, the likelihood of the observed data will always increase, and we obtain the MLEs at the convergence \citep{CPW14}.

\subsection{E step}
In the E step, the expected log likelihood function for the complete data given the observed data and the current parameter estimates is given by
\begin{eqnarray*} 
Q(\bO|\Ot) &=&  \E_{\ymis, \bmm{b} | \yobs,{\bf M}; \Ot} \left[\log L(\bO; \yobs, \ymis, \bmm{b}, {\bf M})\right]. \nonumber \\
&=& \sum_{i\in {\bf O}} \E_{\bi|\bmm{y}_i,M_i;{\bf\Omega}^{(t)} }\ell(\bmm{y}_i, \bi, M_i=0;\bO)+\sum_{i\notin {\bf O}} \E_{ \bmm{y}_i ,\bi |M_i ;\Ot}\ell(\bmm{y}_i, \bi, M_i=1;\bO ) \nonumber \\
&=& I1+I2,
\end{eqnarray*}
where  ${\bf O}$ denotes the set of indices of the observed batches. Existing literature on modelling the non-ignorable missingness in iTRAQ data \citep{Hill08,Luo09,Oberg08} and the selection model for longitudinal data with non-ignorable missingness \citep{Ibrahim90} consider the probability of missingness of a feature in each sample independently. Those methods can handle sample-level non-ignorable missing data, but are not directly applicable to iTRAQ data with batch-level missingness. In contrast, we will take the missing batches into account by explicitly modeling the BADMM -- a major innovation of the proposed method.\\

For the observed batches,
\begin{eqnarray*} 
I1 &=&\sum_{i\in {\bf O}}\E_{\bi|\bmm{y}_i,M_i;{\bf\Omega}^{(t)} } \left\{\log \left[ f(\bmm{y}_i|\ba,\bri,\bi)\right] +  \log \left[ f(\bi|{\bf D})\right] + \log \left[ f(M_i=0|\bmm{y}_i)\right]\right\}.
\end{eqnarray*}
The last term $\log \left[ f(M_i=0|\bmm{y}_i)\right]$ does not involve parameters of interest.

To obtain the conditional expectation, we first calculate the conditional distribution of $\bi$ for $i\in {\bf O}$ as a normal distribution with mean and variance
\begin{eqnarray}\label{eq:omu}
\bmm{b}^{(t)}_{i}=\E(\bi|\bmm{y}_i, M_i=0, \Ot) = {\bf D}^{(t)}\bzi^T\bwi (\bmm{y}_i- \bxi \ba^{(t)}), \\
{\bf\Delta}^{(t)}_i=\var(\bmm{b}_i|\bmm{y}_i, M_i=0, \Ot) = {\bf D}^{(t)}- {\bf D}^{(t)}\bzi^T\bwi \bzi {\bf D}^{(t)}, \label{eq:ovar}
\end{eqnarray}
where $\bwi=({\bf \Sigma}_i^{(t)})^{-1}= (\bzi{\bf D^{(t)}}\bzi^T+{\bf R}_i^{(t)})^{-1}$.
It follows that
\begin{eqnarray*} 
I1 &=& const -1/2 \sum_{i\in {\bf O}} \left( \log|\bri| + {(\bmm{y}_i-\bxi\ba-\bzi\bi^{(t)})}^T\bri^{-1}(\bmm{y}_i-\bxi\ba-\bzi\bi^{(t)})\right. \nonumber \\
&& \left. +  \textrm{tr}({\bf V}_i^{(t)} \bri^{-1} ) + \log |{\bf D}|+{\bi^{(t)}}^T{\bf D}^{-1}\bi^{(t)} + \textrm{tr}({\bf D}^{-1}{\bf \Delta}^{(t)}_i)\right),
\end{eqnarray*}
where ${\bf V}_i^{(t)}=\var(\bmm{e}_i|\bmm{y}_i, M_i=0, \Ot)=\bzi{\bf \Delta}^{(t)}\bzi $ for $i\in {\bf O}$. 

To calculate $I2$, we first compute the conditional expectation and variance of $\bmm{y}_i$ and $\bmm{b}_i$ for $i\notin {\bf O}$. Given $\pr(M_i=1|\bmm{y}_i)$ in equation~(\ref{eq:02}), it is easy to see that, for $i\notin {\bf O}$,
\begin{eqnarray} \label{eq:mmu}
&&\bmm{y}_i^{(t)}=\E(\bmm{y}_i|M_i=1,\Ot)=\bxi\ba^{(t)}-\gamma/p_i{\bf\Sigma}_i^{(t)}\bmm{1},\\
&&\var(\bmm{y}_i|M_i=1,\Ot)={\bf \Sigma}^{(t)}_i, \label{eq:mvar}
\end{eqnarray}

where ${\bf \Sigma}^{(t)}_i=\bzi{\bf D^{(t)}}\bzi^T+{\bf R}_i^{(t)}$. It follows that, for $i\notin {\bf O}$,
\begin{eqnarray}
\bi^{(t)} &  = &\E(\bmm{b}_i| M_i=1, \Ot) = \E(E(\bmm{b}_i|\bmm{y}_i, \Ot)| M_i=1, \Ot) \label{eq:e.bi}\\
        &= &  {\bf D}^{(t)}\bzi^T\bwi (\bmm{y}_i^{(t)}- \bxi \ba^{(t)}),\nonumber
\end{eqnarray}
\begin{eqnarray}
&& {\bf\Delta}^{(t)}_i = \var(\bmm{b}_i| M_i=1, \Ot) \label{eq:var.bi}\\
    &&= \E(\var(\bmm{b}_i|\bmm{y}_i, \Ot)|M_i=1, \Ot)+\var(\E(\bmm{b}_i|\bmm{y}_i, \Ot)|M_i=1, \Ot)\nonumber\\
    &&= {\bf D}^{(t)}, \nonumber\\
 &&{\bf V}_i^{(t)}=\var(\bmm{e}_i|M_i=1, \Ot)=\bri^{(t)}.\label{eq:var.ei}
\end{eqnarray}
Then we can obtain the following for the missing batches of samples
\begin{eqnarray*} 
I2 &=& \sum_{i\notin {\bf O}}\E_{\bmm{y}_i,\bi|M_i;{\bf\Omega}^{(t)} }   \left\{\log \left[ f(\bmm{y}_i|\ba,\bri,\bi,M_i=1)\right] +  \log \left[ f(\bi|{\bf D})\right] + \log \left[ f(M_i=1 |\bmm{y}_i)\right]\right\}  \nonumber \\
&=& const - 1/2 \sum_{i\notin {\bf O}}\left(\log|\bri| +{(\bmm{y}_i^{(t)}-\bxi\ba-\bzi\bi^{(t)})}^T\bri^{-1}{(\bmm{y}_i^{(t)}-\bxi\ba-\bzi\bi^{(t)})} + \textrm{tr}({\bf V}_i^{(t)}\bri^{-1} ) \right.\nonumber \\
&& \hspace{0.5in}\left. +\log |{\bf D}|+{\bi^{(t)}}^T{\bf D}^{-1}\bi^{(t)}+\textrm{tr}({\bf D}^{(-1)}{\bf D}^{(t)}) + 2\gamma/p_i\cdot \bmm{1}^T\bmm{y}_i^{(t)} \right).
\end{eqnarray*}


\subsection{CM step}
In the CM step, we sequentially maximize the expected complete-data log likelihood for the parameters of interest.
In the first step of CM, we obtain the estimate for $\bf D$ that maximizes $Q(\bO | \bO^{(t)})$:
\begin{eqnarray} \label{eq:d}
{\bf D}^{(t+1)} &=& \frac{1}{Q}\sum_{i=1}^Q \left( \bi^{(t)}{\bi^{(t)}}^T+ {\bf\Delta}^{(t)}_i \right).
\end{eqnarray}
Then conditioned on the current $\bri^{(t)}$, the maximum estimate for $\ba$ is given by
\begin{eqnarray} \label{eq:alpha}
&&\ba^{(t+1)}= \left(\sum_{i=1}^Q \bxi^T{(\bri^{(t)})}^{-1}\bxi\right)^{-1}\left(\sum_{i=1}^Q \bxi^T{(\bri^{(t)})}^{-1}(\bmm{y}_i^{(t)}-\bzi\bi^{(t)})\right),
\end{eqnarray}
where $\bmm{y}_i^{(t)} = \bmm{y}_i$ when $M_i=0$, and  $\bmm{y}_i^{(t)} =\bxi\ba^{(t)}-\gamma/p_i {\bf \Sigma}_i^{(t)}\bmm{1}$ when $M_i=1$.

Lastly, we can obtain the estimates for $\sigma^{2(t+1)}_0$ and $\sigma^{2(t+1)}$ conditioned on $\ba^{(t+1)}$:
\begin{eqnarray}\label{eq:sig0}
&{\sigma_0}^{2(t+1)}& = \frac{1}{Q} \sum_{i=1}^Q \left[\left({y}_{i1}^{(t)}-{\bf X}_{i1}\ba^{(t+1)}-{\bf Z}_{i1}\bi^{(t)}\right)^2 + v^{(t)}_{i11}\right], \textrm{ and }\\
\label{eq:sig}
& {\sigma}^{2(t+1)}& =  \left\{\sum_{i=1}^Q \left[\sum_{j=2}^{p_i}\left({y}_{ij}^{(t)}-{\bf X}_{ij}\ba^{(t+1)}-{\bf Z}_{ij}\bi^{(t)}\right)^2 \right.\right.  \\
&& \hspace{1in} + \left .\left.\left(\textrm{tr } {\bf V}^{(t)}_{i} - v^{(t)}_{i11}\right)\right]\right\}/(\sum_{i=1}^Q p_{i}-Q), \nonumber
\end{eqnarray}
where $v^{(t)}_{i11}$ denotes the first diagonal element of ${\bf V}^{(t)}_{i}$.
By iterating through the E- and CM- steps, MLEs for the fixed effects and variance components can be obtained.

In addition, through computing the information matrix of the log-likelihood function of the observed data, we can estimate the variance of $\hat\ba$ using
\begin{eqnarray} \label{eq:var.alpha}
\widehat{\var}(\hat\ba)= \left(\sum_{i \in {\bf O}} \bxi{\bf W}_i\bxi \right)^{-1}.
\end{eqnarray}We can then perform test for $\ba$ by rejecting $H_0:\ \alpha_i=0$ for large value of $\hat\alpha_i/\widehat{sd}_{\hat\alpha_i}$.

\subsection{Estimation of the missing-data mechanism parameter}

In real applications, the missing-data mechanism parameter ${\bf \Gamma} = \{\gamma_0, \gamma\}$ in (\ref{eq:02}) 
is often unknown and needs to be estimated. One simple approach is to use the missing percentage and mean abundance based on available data of each feature to model the relationship between the probability of missingness and the abundance. Specifically, we assume all of the features in one data set are subject to the same missing-data mechanism.
We calculate the average abundance for each feature $j$ based on the observed data and denote it as $t_j$, and we also obtain the missing percentage of feature $j$ as $\pi_j=1-Q_{j,obs}/Q$, where $Q_{j,obs}$ is the number of batches in which feature $j$ is quantified. We can estimate ${\bf\Gamma}$ in (\ref{eq:02}) by
\begin{eqnarray} \label{eq:est_gamma}
{\bf\hat\Gamma} =\Argmin_{\bf\Gamma=\{\gamma_0,\gamma\}} \sum_j \left(\textrm{log}(\pi_j) + \gamma_0 + \gamma  t_j \right)^2.
\end{eqnarray}

Alternatively, one can also employ a profile likelihood approach proposed in \citet{CPW14} to jointly estimate the parameters of interest and the missing-data mechanism parameters.  Let $L_{\bf\Gamma}({\bf\Omega})=L(\yobs,{\bf M};{\bf \Omega},{\bf \Gamma}).$ One can evaluate $L_{\bf\Gamma}({\bf\Omega})$ at different $\bf\Gamma$ values and choose the $\bf\Gamma$ that gives the maximum over the likelihood profile. As shown in \citet{CPW14} with both simulations and real data examples, the estimated ${\bf \Gamma}$  based on available-case estimates of protein abundance is very close to the profile-likelihood estimates, especially when the sample size is limited as in most proteomics studies. Moreover, in Section 5.3, we demonstrate that the available-case estimate of $\bf\Gamma$ is very close to the true values under all the simulation settings considered in this paper. Thus, we use the available-case estimates of the missing-data mechanism parameter in our data analysis.

\subsection{An outline of the algorithm to fit the mixEMM model}
In summary, we implement an ECM algorithm to fit the \texttt{mixEMM} model for analyzing iTRAQ proteomics data. An outline of the \texttt{ECM} algorithm is provided in Algorithm~\ref{alg:mixEMM}.
\begin{algorithm}
\caption{An algorithm to fit the mixEMM model.} \label{alg:mixEMM}
\begin{enumerate}[1.]
\item Estimate missing-data mechanism parameter $\bf \Gamma$ by (\ref{eq:est_gamma}).
\item Obtain the initial estimate ${\bf \Omega}^{(0)}$ for fixed effects and variance components.
\item E-step: For the exponential missing-data mechanism function, given $\bf \hat\Gamma$, calculate the conditional expectations and variances of $\ymis, \bmm{e}_i,\bmm{b}_i$ given the observed $\yobs$, ${\bf M}$ and the current parameter estimates ${\bf\hat \Omega}^{(t-1)}$, according to (\ref{eq:omu}), (\ref{eq:ovar}), (\ref{eq:mmu})  and (\ref{eq:mvar}).
\item CM-step: Given the estimated sufficient statistics, obtain the current estimates of $\bf D$, $\bmm{\alpha}$, $\sigma_0^2$ and $\sigma^2$, using (\ref{eq:d}), (\ref{eq:alpha}) and (\ref{eq:sig0}) and (\ref{eq:sig}), respectively.
\item Repeat 3-4 until convergence.
\end{enumerate}
\end{algorithm}
Note, for the small amount of sporadic missingness, we treat them as MAR and remove the corresponding data points from the evaluation of the likelihood function. Specifically, if a protein is measured in $l\ (l<4)$ samples in a 4-plex iTRAQ experiment, we will set $p_i=l$ and apply the proposed method.

\subsection{Logit probability functions for BADMM}
The probability of missingness in (\ref{eq:02}) is designed to characterize the BADMM for abundance data from iTRAQ or other proteomics experiments.
By using an exponential function, the probability of missingness in (\ref{eq:02}) can be naturally integrated with the density function of normal distributions. Thus, closed-form solutions can be obtained in the ECM algorithm, which makes the computation efficient.

In some instances, for example, when tumor and normal samples are matched and paired, and each pair is considered as an ``observation" in the iTRAQ experiment, there may be a need to analyze log-ratio data (i.e. log of ratios of abundances of a tumor sample versus the matching normal sample). In this situation, if a feature has low abundance in either the tumor or the normal sample, the observed log-ratio values for the pair would be extreme, and the feature would be more likely to be missing. The exponential function-based BADMM is not suitable to model the missing-data pattern for log-ratio  data.
In those cases, we could use a more general and flexible logistic function for modelling the missing-data mechanism \citep{LR02,Luo09}:
\begin{equation} \label{eq:logit}
\textrm{logit}(\pr(M_i=1|\bmm{y}_i))=  \gamma_0+\gamma /p_i \cdot \bmm{1}^T \bmm{y}_i + \bmm{\gamma}_2\cdot {\bf C}_i,
\end{equation}
where ${\bf C}_i$ is a set of covariates associated with the experiment $i$ (or the $i$-th batch), and $\bmm{\gamma}_2$ is the corresponding coefficient. In our motivating example, we do not have any experiment-specific (nor batch-specific) covariate, and thus the last term is not considered.

For the logit missing-data mechanism in (\ref{eq:logit}), we will use numeric integration \citep{PB95} to obtain the conditional means and variances for $\bmm{y}_i^{(t)}$'s in the missing batches, and replace the corresponding terms in (\ref{eq:e.bi}), (\ref{eq:var.bi}) and (\ref{eq:var.ei}) with the following:
\begin{eqnarray} 
&&\bmm{y}_i^{(t)}=\E(\bmm{y}_i|M_i=1,\Ot)=  \frac{\int{\bmm{y}_i P(M_i=1|\bmm{y}_i)\phi(\bmm{y}_i, \bxi\ba^{(t)}, {\bf \Sigma}_i^{(t)}) d\bmm{y}_i }}{ \int{ P(M_i=1|\bmm{y}_i)\phi(\bmm{y}_i, \bxi\ba^{(t)}, {\bf \Sigma}_i^{(t)}) d\bmm{y}_i } }, \nonumber \\
&&\var(\bmm{y}_i|M_i=1,\Ot)= \E(\bmm{y}_i\bmm{y}_i^T|M_i=1,\Ot)-\bmm{y}_i^{(t)} {\bmm{y}_i^{(t)}}^T; \nonumber  \\
&&\bi^{(t)}=\E(\bmm{b}_i|\bmm{y}_i, M_i=1, \Ot) = {\bf  D}^{(t)} \bzi^T\bwi\left(\bmm{y}_i^{(t)} - \bxi\ba^{(t)} \right),  \nonumber  \\
&& \var(\bmm{b}_i|\bmm{y}_i, M_i=1, \Ot) = {\bf D}^{(t)} - {\bf  D}^{(t)} \bzi^T\bwi \bzi{\bf  D}^{(t)} + \nonumber  \\
&& \hspace{2in} {\bf  D}^{(t)} \bzi^T\bwi \var(\bmm{y}_i|M_i=1,\Ot) \bwi\bzi{\bf  D}^{(t)};\nonumber  \\
&& \textrm{and } {\bf V}^{(t)}_i = \var(\bmm{e}_i|\bmm{y}_i, M_i=1, \Ot)= \bzi{\bf D}^{(t)}\bzi^T - \bzi{\bf  D}^{(t)} \bzi^T\bwi \bzi{\bf  D}^{(t)}\bzi^T \nonumber  \\
&& \hspace{2.5in} + \bri\bwi \var(\bmm{y}_i|M_i=1,\Ot) \bwi \bri .\nonumber
\end{eqnarray}

\section{Simulations}
\subsection{Comparison of modelling absolute abundance via mixed-effects models versus modelling relative abundance}
To remove batch-effects in iTRAQ-based proteomics analyses, a standard practice is to analyze the relative abundance of a protein/peptide in the target samples relative to the abundance level of the protein/peptide in the reference sample from the same batch, and assess the association of relative abundance of each protein/peptide with the phenotype. In this simulation section, we will show that directly modelling absolute abundance with mixed-effects model-based approaches improve over conventional analysis based on relative abundances.

We simulated 1,000 multivariate normal data sets $\bmm{y}_i \sim N(\bxi \ba + \bzi\bmm{b}_i, {\bf R})$ with $p=4$ for batch size of $Q=40$ and $Q=200$.
The fixed effects are $\bxi \ba$. Here $\bxi$ is a $p \times (k+1)$ ($k=2$) covariate matrix for each observation $i$ with the first column being $\bmm{1}$, and $\ba=(10, -a , a)^T$. In assessing the type I error rate, we set $a=0$; in evaluating the power, we set $a=0.7$ when $Q=40$ and $a=0.3$ when $Q=200$. Here we only included a random intercept for each batch. The random effect is $b_i\sim N(0, D)$, and $\bzi$ is a vector of 1's. $\bf{R}$ is a diagonal matrix with diagonal elements $(\sigma_0^2,\sigma^2, \sigma^2, \sigma^2)$. We simulated two scenarios: when the experimental variation is large, $\sigma_0^2=2,\sigma^2=4,D=3$; and we also simulated a scenario with smaller experimental variation,  $\sigma_0^2=1,\sigma^2=2,D=1$.  Note $\sigma_0^2$ represents the variance of the reference sample,
which is purely due to experimental variation across different iTRAQ multiplex; while $\sigma^2$ represents the variance of the target samples, which is a combination of both biological and experimental variation. Thus, the reference sample variance is often smaller than the variance of other tumor samples. We generated approximately 40\% missing data at the batch-level by the mechanism in (\ref{eq:02}) with $\gamma_0=0$ and $\gamma=0.1$
We also generated an additional 5\% sporadic (random) missingness.

When applying the \texttt{mixEMM} method, based on the estimated MLEs for the fixed effects and their variance estimates in (\ref{eq:var.alpha}), we first obtained the Wald test statistics for testing $H_0:\bmm{\alpha}_{-1}=\bmm{0}$, where $\bmm{\alpha}_{-1}$ stands for the fixed effects other than the mean (i.e. the intercept), and then derived the $p$-values by approximating the null distribution through permuting the order of batches of response variables. We compared two versions of the \texttt{mixEMM} method: one with and one without incorporating BADMM by setting $\gamma=0.1$ and $0$, respectively. Note, when $\gamma=0$, the missing mechanism is treated as MAR (missing at random).

We also compared the performance of \texttt{mixEMM} with that of the conventional analysis based on relative abundances: we treated relative abundances as responses and fitted linear regressions to detect significant associations (regression coefficients). Again, $p$-values were derived through permutation tests in the same way as we did for \texttt{mixEMM}.

Table~\ref{table:pwr} shows that with permutation-based $p$-values, all the three methods can control  type I error rates at different $p$-value thresholds  in different scenarios. Comparing with the conventional approach of analyzing relative abundance, both versions of \texttt{mixEMM} enjoyed much improved power. In particular, when experimental variation is large, the improvement of power could be 3 to 4 folds. These results clearly demonstrated the advantage of modeling the batch design through a mixed-effects model, which helps to characterize the variance structure in the data more precisely. The two versions of \texttt{mixEMM} ($\gamma=0.1$ v.s. $\gamma=0$) enjoys similar power in all settings. This suggests that BADMM has only limited impacts on the testing results.  However, in the next section, we demonstrated that incorporating BADMM will improve parameter estimation.

 \begin{table}
\caption{Type I error rate and power comparison. We compare the type I error rates and power of the \texttt{mixEMM} method with and without considering BADMM, as well as linear regressions using relative abundances as responses.} \label{table:pwr}
\begin{center}
\begin{adjustbox}{width=1\textwidth}
\begin{tabular}{|c|c|c|c|c|c|c|}\hline
&\# batch    &           & $P$-value &\texttt{mixEMM}    & \texttt{mixEMM} & Linear regression \\
&(experiment)&  Variance & cutoff    & $\gamma=0.1$    &  $\gamma=0$       &on relative abundance  \\ \hline
  &  \multirow{4}{*}{$40$} & large  &0.05 &   0.055 & 0.056  & 0.048   \\ \cline{4-7}
    &                      &  & 0.01 &  0.007 &  0.007 &  0.014\\ \cline{3-7}
     &                    &  small  &0.05 &  0.060 & 0.063  &  0.047  \\ \cline{4-7}
  Type I  &              &     & 0.01 &  0.012 &  0.012 &  0.012 \\ \cline{2-7}
error & \multirow{4}{*}{$200$}& large  &0.05 &   0.045 & 0.045  & 0.058  \\ \cline{4-7}
 rate    &                 &  & 0.01 &  0.008 & 0.008  &  0.007\\ \cline{3-7}
         &                 &  small  &0.05 &  0.055  & 0.057  &  0.050  \\ \cline{4-7}
         &                &     & 0.01 &  0.008 & 0.010  & 0.017 \\ \cline{2-7}
\hline
  &  \multirow{4}{*}{$40$} & large  &0.05 &   0.437 & 0.442  & 0.150   \\ \cline{4-7}
    &                      &  & 0.01 &  0.267 &  0.263 &  0.068\\ \cline{3-7}
     &                    &  small  &0.05 &  0.959 & 0.957  &  0.507  \\ \cline{4-7}
   &              &     & 0.01 &  0.898 &  0.899 &  0.240 \\ \cline{2-7}
Power & \multirow{4}{*}{$200$}& large  &0.05 &   0.491 & 0.472  & 0.178  \\ \cline{4-7}
     &                 &  & 0.01 &  0.248 & 0.259  &  0.065\\ \cline{3-7}
         &                 &  small  &0.05 &  0.979  & 0.979  &  0.555  \\ \cline{4-7}
         &                &     & 0.01 &  0.895 & 0.901  & 0.306 \\ \cline{2-7}
\hline
\end{tabular}
\end{adjustbox}
\end{center}
\end{table}

\subsection{The BADMM modeling in \texttt{mixEMM}}
We simulated 1,000 multivariate normal data sets similar as before with $\ba=(10, -1 , 1)^T$, $\sigma_0^2=2$, $\sigma^2=4$.  We generated approximately 40\% missing data at the batch-level by the mechanism in (\ref{eq:02}) with $\gamma_0=0$ and $\gamma=0.1$, and  an additional 5\% sporadic (random) missingness.

Table~\ref{table:01} shows the relative Mean Squared Errors (MSEs) of \texttt{mixEMM} incorporating BADMM ($\gamma=0.1$) versus \texttt{mixEMM} without considering BADMM ($\gamma=0$) on estimates for the fixed effects and variance with different sample sizes. The relative MSEs for the fixed effects estimates are approximately 0.8 for $Q=40$, and 0.5 for $Q=200$. This suggests that by taking into account the missing batches, the proposed \texttt{mixEMM} method provides more accurate estimates for fixed effects in both the limited and large sample scenarios. The relative MSEs for variance estimates are very close to 1, indicating that modelling the non-ignorable missingness mainly helps to correct the biases in the fixed effects estimates rather than variance estimates.

In addition to the simulations above, we also re-analyzed the simulated data using the logit missing-data mechanism function in (\ref{eq:logit}) and compared the relative MSEs. Note that the simulated data are generated from the exponential BADMM in (\ref{eq:02}), and we use the logit function to analyze the data, with $\gamma_0=0$ and $\gamma=0.1$. That is, the missing-data mechanism is potentially mis-specified. The relative MSEs based on the logit function are close to those based on the true BADMM, with only a minor loss of efficiency. Since the logit function is quite flexible and fits the observed missing-data pattern well,  the overall biases of the fixed effects estimates are quite small. This suggests that the logit BADMM function is a general and flexible missing-data mechanism function.  When data are generated by logit BADMM and re-analyzed by exponential BADMM, as long as the exponential pattern nicely fits the observed missing data pattern, the conclusions are similar (results not shown).

When the two BADMM mechanisms produce similar patterns, the exponential function is about 15 times faster than the logit function.  Specifically, in terms of computation time, it takes 0.287 and 1.514 hours for a single node computer to analyze 1,000 features based on the exponential BADMM when sample sizes are 40 and 200, respectively, whereas it takes 4.460 and 24.243 hours for the analysis based on the logit BADMM.  The computation time increases rapidly with dimensionality $p$ and sample size $n$. When jointly analyzing multiple features, for example multiple peptides from iTRAQ data with 8 channels, the superiority of the exponential BADMM would become more substantial. Additionally, the logit BADMM would be useful in analyzing log-ratio data or when exponential pattern does not fit well.

The fit of the selected and estimated BADMM pattern should often be checked before using the \texttt{mixEMM} method in the estimation and inference. For example, in our real data application, we evaluate the fit of the exponential BADMM in Figure~\ref{fig:BADMMexp} before the subsequent analysis.

\begin{table}
\caption{The comparison of relative MSEs and computation time for estimates of fixed effects and variance components obtained from incorporating BADMM ($\gamma=0.1$) relative to those assuming MAR ($\gamma=0$) in \texttt{mixEMM}.  The missing data are generated according to the exponential BADMM in (\ref{eq:02}). We compare the relative MSEs when the true missing-data mechanism is accounted in the estimation of the \texttt{mixEMM} algorithm, and when the logit BADMM is used in the estimation with estimated missing-data mechanism parameters.  The results are based on 1,000 repeated simulations.}\label{table:01}
\begin{center}
\begin{tabular}[i]{|c|c|c|c|c|c|c|}\hline
           &&&&&&Computation\\
 Methods &  \# experiment &  $\ba$ & $\sigma_0^2$ & $\sigma^2$ & $D$ &  time  \\
         &  $Q$ &        &             &             &     &   (in hours)        \\ \hline
\texttt{mixEMM} with &    40     &    0.848  &  1.014 &  1.006 &  1.184 & 0.287      \\ \cline{2-7}
 exponential BADMM &    200       &     0.492 & 1.016 & 1.006  & 1.015 & 1.514  \\ \hline
\texttt{mixEMM} with   & 40 &       0.851 & 1.004  & 1.002 & 1.047 &  4.460        \\ \cline{2-7}
logit BADMM & 200 &     0.538 &  1.007 & 1.004 & 0.983 &  24.243    \\ \hline
\end{tabular}
\end{center}
\end{table}

\subsection{Evaluating available-case-based missing-data mechanism parameter estimates}
In the simulations above when applying the \texttt{mixEMM}, we either used the true missing-data mechanism with true parameters, or we use a mis-specified mechanism with mis-specified parameters. In this subsection, we evaluated the estimation of the missing-data mechanism parameter. Specifically, we pooled all features together and obtained the available-case mean abundance estimates and the proportion of missing batch for each feature. We estimated the missing-data mechanism parameter based on (\ref{eq:est_gamma}) for the exponential BADMM in (\ref{eq:02}) given the data.

We simulated 1,000 features with means randomly sampled from $N(10,2^2)$, other parameters similar to those in previous sections, and the number of batches $Q=40$ and $Q=200$. We generated batch-level missingness by (\ref{eq:02}) with $\gamma_0=0$ and $\gamma=0.1$ and calculated $\hat\gamma_0$ and $\hat\gamma$ based on the 1,000 features.  We repeated the procedure 100 times, and Table~\ref{table:03} lists the distribution of $\hat\gamma_0$ and $\hat\gamma$.  Although available-case based mean estimates of feature abundances could have substantial biases, the estimates for $\gamma$ are reasonably accurate. We note that estimates for $\gamma_0$ based on available-case means can be biased. However, since $\gamma_0$ does not affect the E- nor  the CM-step, the overall performance of available-case-based missing-data mechanism parameters is almost identical to that of using true parameters.

 \begin{table}
\caption{The distribution of available-case-based estimated missing-data mechanism parameters based on 100 repeated simulations.}\label{table:03}
\begin{center}
\begin{tabular}{|c|c|c|c|c|c|c|}\hline
 \# batch  & parameter  & true value & min &  median & mean & max \\ \hline
 \multirow{2}{*}{40} &  $\gamma$ & 0.1 & 0.093 & 0.101 & 0.101 & 0.107  \\\cline{2-7}
      &  $\gamma_0$ &  0  & -0.119 & -0.059 & - 0.055 & 0.029 \\ \hline
 \multirow{2}{*}{200} &  $\gamma$ &  0.1   &  0.097    &  0.104  & 0.104    & 0.108 \\ \cline{2-7}
         &  $\gamma_0$ &  0  &  -0.134  & -0.094  & -0.093   &  -0.014 \\
\hline
\end{tabular}
\end{center}
\end{table}

\section{Application to the CPTAC proteomics data to identify proteins related to triple negative breast cancer tumors}
 Triple negative breast cancer (TNBC) refers to breast cancer that does not express the genes for estrogen receptor, progesterone receptor or Her2/neu. TNBC patients have a much higher risk of relapse for the first 3–5 years compared to other types of breast cancer patients. It is also more difficult to treat TNBC since most chemotherapies target one of the three receptors. There is a pressing need to develop more effective treatment strategies for TNBC patients. In this section, we applied the proposed \texttt{mixEMM} algorithm to the motivating proteomics dataset from the CPTAC breast cancer project \citep{CPTAC,CPTAC2}, with the goal of identifying phosphopeptides up or down regulated in TNBC tumors compared to other types of breast cancer tumors. Such information can help to shed light on the disease mechanism of TNBC, which then may lead to better clinical practice for TNBC diagnosis and treatment.

In the CPTAC breast cancer project, a total of 108 tumor samples from 105 breast cancer patients were analyzed in 36 four-plex iTRAQ experiments generated at Dr. Carr's lab at the Broad Institute of MIT and Harvard, Boston, US. Each iTRAQ experiment processed 3 breast tumor samples and the reference sample, which was created by combining 40 of these tumors. The iTRAQ-labeled peptides were fractionated and chemically enriched for phosphopeptides. The resulting samples were processed using high resolution MS instruments (LS-MS/MS on Thermo Q-Exactive). Phosphopeptide identification and quantification were performed using Spectrum Mill software (Agilent Technologies, Santa Clara, CA).

In this application, we focus on phospho-proteomics data and will analyze each individual phosphopeptide. Missing data problem in phospho-proteomics data is usually quite severe, and thus raises a pressing need for statistical methods properly incorporating the non-ignorable missingness.
In total, there were 61,698 phospho-peptides being identified and quantified in at least one sample. However, only 4,415 ($7.2\%$) phosphopeptides had complete measurements in all the samples. The missing rates of each sample ranged from $58.21\%$ to $83.40\%$. Among all missing observations, $99.3\%$ were batch-level missingness, i.e. a phosphopeptide was missing in all four channels of an iTRAQ experiment. Thus the BADMM pattern suits these data sets well.

\begin{figure}[!h]
\begin{center}
\includegraphics[scale=0.4]{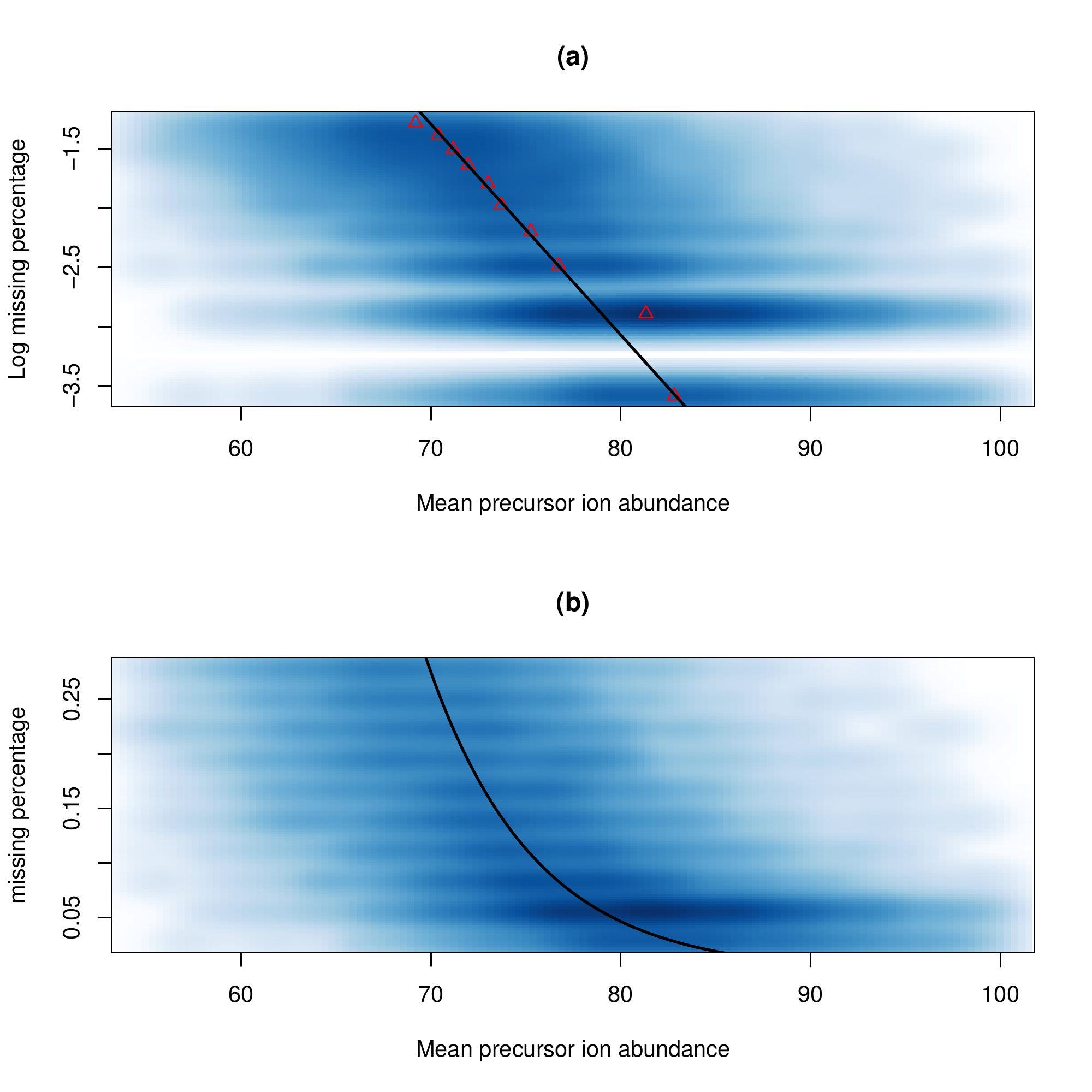}
\end{center}
\caption{An illustration of BADMM based on CPTAC breast cancer phosphoproteomics data. (a)  A smoothed color density representation of the scatter plot of the log percentage of missing batches for each phosphopeptide (y-axis) versus its estimated mean abundances based on the observed data (x-axis). This plot is generated using R function \texttt{smoothScatter}. The darker the color is, the higher the density is. The red triangular points indicate medians of mean-abundances of phosphopeptides with the same missing percentage. The black line represents the linear regression fit of the red triangular points. (b) Similar plot as (a) except that the y-axis is on the original scale. The black curve corresponds to the black line in (a). }\label{fig:BADMMexp}
\end{figure}

We filtered out the low quality observations, and focused on the 25,961 phosphopeptides that were observed in at least 25 ($70\%$) of the 36 runs of the reference sample. The missing rates of each sample for these 25,961 phosphopeptides ranged from  $3.98\%$  to $48.44\%$ , with a mean value of $10.45\%$. Figure~\ref{fig:BADMMexp}  illustrates the relationships between missing percentage and observed mean abundances of each phosphopeptide (i.e. $\mathbf{1}^T \mathbf{y}_i$ in equation (\ref{eq:02})). The exponential probabilistic model in equation (\ref{eq:02}) accurately reflects the BADMM pattern in the data.

\begin{figure}[!h]
\begin{subfigure}{0.44\textwidth}
\includegraphics[width=\linewidth,angle=270]{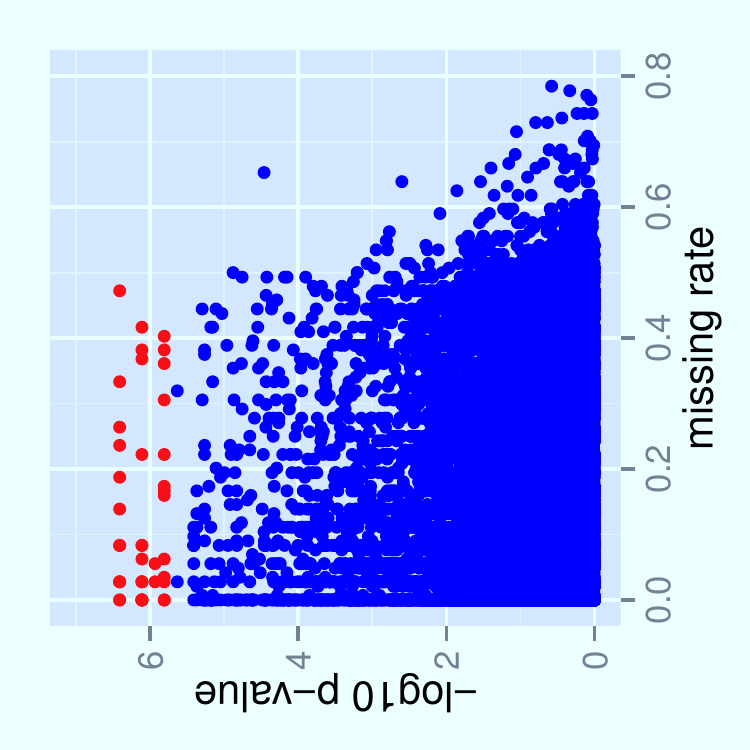}
\caption{mixEMM} \label{fig:pvalue.a}
\end{subfigure}
\hspace*{\fill} 
\begin{subfigure}{0.44\textwidth}
\includegraphics[width=\linewidth,angle=270]{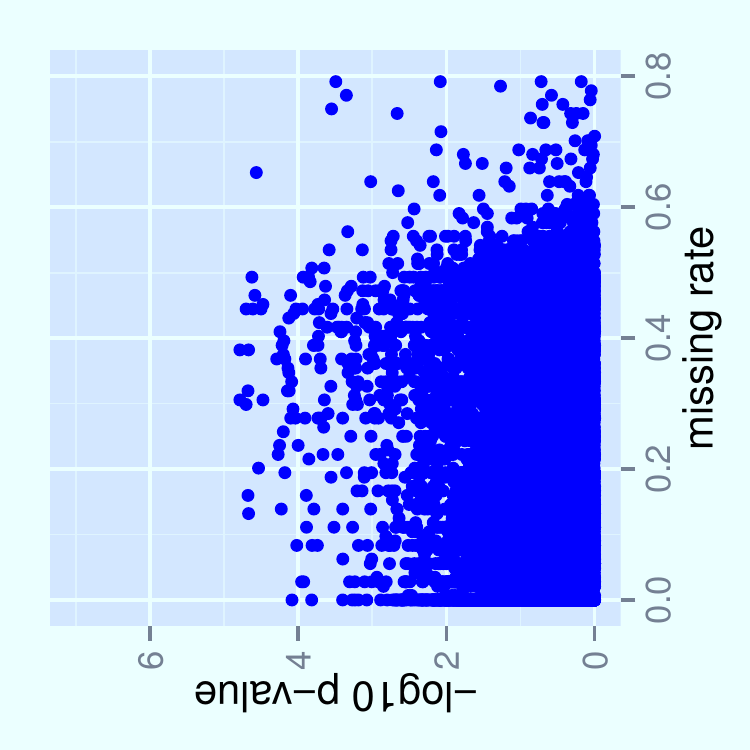}
\caption{Linear Regression} \label{fig:pvalue.b}
\end{subfigure}
\caption{The relationship between $p$-values and missing rates. (a) The results from \texttt{mixEMM}-based analysis using absolute abundances and considering BADMM; (b) the results from linear regression analysis using relative abundances. In both (a) and (b), X-axis represents the missing rates of phosphosites and Y-axis represents the negative $log_{10}$ of $p$-values. Phosphopeptides are colored in red if their $p$-values are below the Bonferroni corrected $p$-value cutoff ($0.05/25961$), and are colored in blue otherwise.  }\label{fig:pvalue}
\end{figure}

We applied the proposed \texttt{mixEMM} method to identify the phosphopeptides up- or down-regulated in TNBC tumors. In the mixed-effects model, we included a random effect for each iTRAQ multiplex experiment, and three fixed effects: an intercept, an indicator for the reference channel, and an indicator for triple negative subtype. We also conducted the analysis using linear regression models based on relative abundances for comparison. The resulting $p$-values of all 25,961 phosphopeptides from both methods are illustrated in Figure~\ref{fig:pvalue}. At Bonferroni adjusted $p$-value threshold of 0.05, the \texttt{mixEMM} algorithm considering BADMM identifies 44 phosphosites, corresponding to 29 unique genes, as being significantly up or down-regulated in TNBC. Only 3 of these 44 phosphosites have complete observations in all 108 samples. Nine and three of the 44 phosphosites have a missing rate great than $30\%$ and $40\%$ respectively. In contrast, the conventional analysis based on relative abundances failed to detect any significant phosphosite at the same significance threshold. These results are consistent with what we observed in the simulated data examples, and the \texttt{mixEMM} method enjoys improved power over conventional methods.

 The phosphosite with the most significant $p$-value corresponds to the gene {\it FOXA1}, a transcription factor. The gene {\it FOXA1} is known to be associated with breast cancer risk \citep{Meyer2012}. A more recent work further suggests that FOXA1 silencing increases migration and invasion of breast cancer cells \citep{Bernardo2013}. This is consistent with our finding that phosphoprotein of {\it FOXA1} was significantly down-regulated in TNBC tumors, and TNBC tumors are usually more aggressive than other subtypes of breast cancer. Moreover, according to the STRING data base \citep{szklarczyk2014string}, {\it FOXA1} interacts with another gene {\it SOX10} --- in the significant 29 gene list. The gene {\it SOX10} is a neural crest transcription factor. It was reported to be preferentially expressed in TNBC based on a recent immunohistochemistry study \citep{cimino2013neural}, and was also validated as a sensitive diagnostic marker for basal-like TNBC \citep{ivanov2013diagnostic}. The proposed \texttt{mixEMM} method detects these known TNBC genes and that  strengthens our confidence that the \texttt{mixEMM} method will help to  reveal biological relevant information underlying iTRAQ data. Further investigation on how {\it FOXA1}, {\it SOX10} and the other 27 significant genes function may help us better understand the disease mechanism and improve the development of novel diagnostic and therapeutic tools for TNBC.


\section{Discussion}
In this paper, we propose a new method --- \texttt{mixEMM} --- for analyzing data from iTRAQ proteomics experiments. The proposed \texttt{mixEMM} method  employs a mixed-effects model to characterize the variance structure for the abundance measurements from iTRAQ experiments. It uses an exponential probability function to model the batch-level non-ignorable missing-data mechanism (BADMM) in the iTRAQ data. The goal of our analyses is to estimate the fixed effects for the association between proteomic features and sample phenotypes (e.g. clinical outcomes). To achieve this goal, we implement an ECM algorithm to calculate the MLEs of the parameters of interest. The superior performance of the \texttt{mixEMM} method over the conventional analysis approach is illustrated using both simulations and a real data example.

In practice, the experimental variation across different iTRAQ experiments is often not small. In other words, even for the same reference sample, its protein/peptide abundance measurements in different batches measured by different iTRAQ experiments may differ substantially. The conventional approach directly analyzes relative abundance measures, which in some sense mix up the variation in the target samples and the reference samples, and consequently causes a loss of efficiency and power. In contrast, \texttt{mixEMM} precisely characterizes the experimental properties, accounts for the variation of reference sample across batches, and gain substantial power improvement in the subsequent tests.

While explicitly modeling BADMM has limited impact on testing, it can improve parameter estimations for fixed effects. In addition to the exponential probability function for BADMM, we also investigate the use of logit functions for modelling the missing-data mechanism. When both functions fit the observed missing-data pattern well, the estimation accuracies of the two functions are comparable, and the computationally efficient exponential function is recommended. The logit BADMM function is more flexible and can be used in the analyses of log-ratio data or data with more complex missing data patterns. Other flexible missing-data mechanism functions such as spline functions can be incorporated into the proposed framework, although numerical integration would be required.

This work was motivated by phosphoproteomics data, in which the natural analysis unit is each individual phosphopeptide and each phosphopeptide is directly quantified in the experiments. For global proteomics data, the quantification is obtained at the peptide level while the target analysis unit is each individual protein. In order to perform inference at the protein level, one strategy is to apply the proposed \texttt{mixEMM} algorithm at the peptide level data and then summarize the results of peptides within each protein. Another strategy is to first calculate protein abundances based on the mean or median of peptide abundances within each protein and then apply the proposed \texttt{mixEMM} method to the summary protein abundances. A more sophisticated treatment would be to perform a multivariate analysis and jointly model multiple peptides of the same protein. Research along this direction is on-going. 

The proposed framework is not limited to proteomics data analysis, and is generally applicable to data with repeated/clustered measures and cluster-level incomplete data. An R package \texttt{mixEMM} will be available through CRAN.

\section*{Acknowledgement}
Mass spectrometry and proteomics data were acquired by the breast cancer project of CPTAC consortium, which is led by Dr. Steve Carr from Broad Institute of MIT and Harvard, and Dr. Amenda Paulovich from Fred Hucthinson Cancer Research Center, and is supported by NCI grant CA160034. We thank Drs. Chenwei Lin,  D.R. Mani,  Philipp Mertins,  Yan Ping and others from the CPTAC consortium for their help on the proteomics data. We also thank Drs. Ross Prentice for helpful suggestions and comments. LSC and JW are supported by R01GM108711 and R03CA174984. XW and PW is supported by SUB-CA160034. PW is also supported by NIH grant P01CA53996, SUB-R01GM108711 and R01GM082802.

\bibliographystyle{imsart-nameyear}
\bibliography{mixedBADMM}

\end{document}